# Lie-point symmetries of the Lagrangian system on time scales

Cai Ping-Ping[1], Song-Duan[2], Fu Jing-Li[1*], Fang-Yu Hong[1]

[1]*Institute of Mathematical Physics, Zhejiang Sci-Tech University, Hangzhou, 310018, China*

[2]*Eastern Liaodong University, Shenyang, 118000, China*

**Abstract**

This letter investigates the Lie point symmetries and conserved quantities of the Lagrangian systems on time scales, which unify the Lie symmetries of the two cases for the continuous and the discrete Lagrangian systems. By defining the infinitesimal transformations' generators and using the invariance of differential equations under infinitesimal transformations, the determining equations of the Lie symmetries on time scales are established. Then the structure equations and the form of conserved quantities with delta derivatives are obtained. The letter also gives brief discussion on the Lie symmetries for the discrete systems. Finally, several examples are designed to illustrate these results.

## 1. Introduction

The theory of time scales is a relatively new field, introduced by Stefan Hilger in 1988 [1] in order to unify and generalize difference and differential equations. Time scale calculus theory is applicable to any field in which dynamic processes can be described with discrete or continuous models. The study of the calculus of variations in the context of time scales has its beginning in 2004 with the paper[2] of Martin Bohner. Since the pioneer paper [2], the classical results of the calculus of variations on continuous-time ($\mathbb{T}=\mathbb{R}$) and discrete-time($\mathbb{T}=\mathbb{Z}$) have been unified and generalized to a time scale $\mathbb{T}$: Euler-Lagrange equations [3,4]; necessary optimality conditions for variational problems subject to isoperimetric constraints [5,6]; high-order delta derivatives [7-9]; weak maximum principle for variable endpoints optimal control problems [10]; integration on time scales [11]; the boundary value problems [12,13]; applications of time scale to economics [14]. In recent years, D. F. M. Torres and

*Corresponding author. E-mail:sqfujngli@163.com
Supported by the National Natural Science Foundations of China (No. 11072218) and the National Natural Science Foundations of Zhejiang Province of China (Grant No. Y6110314) others made use of the Euler-Lagrange equations on time scales to generalize one of the most beautiful results of the calculus of variations-the celebrated Noether's theorem [4,15].

It is generally known that the principle of symmetry and the laws of physics have close relations. And the development of modern physics such as quantum mechanics, quantum field theory and nuclear physics shows that the principle of symmetry has become the most important principles of exploring the laws of motion of microparticles. The analysis of the symmetries and conserved quantities of physical systems is very important to study the dynamical behavior of the systems and their qualitative properties. There are two modern methods to find the conserved laws, that is Noether symmetry method and Lie symmetry method. The Noether method is making good progress [16-19]. The Lie group theory has been approved to be a powerful tool to solve differential equations, to study constrained mechanical systems, to discuss controllable dynamical systems, to investigate mechanico-electrical systems and to establish properties of their solution space. These aspects of Lie group theory have been described in many literatures [20-27]. Lie group theory has also been applied to discrete equations, such as differential-difference equations, discrete dynamical systems and discrete mechanico-electrical systems [21,28,29].

In this paper, the Lie method is introduced to explore the symmetries of the Lagrangian systems on an arbitrary time scale $\mathbb{T}$. The determining equations, the structure equations and forms of conserved quantities with delta derivatives of the Lie symmetries are given. The Lie-form invariant (Theorem 3) unifies and extends the previous formulations of Lie's method in the discrete-time and continuous domains [20-29]. The Lie symmetries of the discrete Lagrangian systems are also discussed and several examples to illustrate the application of the results are given.

## 2. Basics on time scale calculus

A time scale is a nonempty closed subset of real numbers, and we usually denote it by the symbol $\mathbb{T}$. The two most popular examples are $\mathbb{T}=\mathbb{R}$ and $\mathbb{T}=\mathbb{Z}$. We define the forward and backward jump operator $\sigma, \rho: \mathbb{T} \to \mathbb{T}$ by

$$\sigma(t)=\inf\{s\in\mathbb{T}:s>t\} \quad \text{and} \quad \rho(t)=\sup\{s\in\mathbb{T}:s<t\}, \text{ for all } t\in\mathbb{T},$$

(supplemented by inf Ø=sup$\mathbb{T}$ and supØ=inf $\mathbb{T}$). The graininess function $\mu: \mathbb{T}\to[0,\infty)$ is defined by

$$\mu(t)=\sigma(t)-t. \tag{1}$$

Hence the graininess function is constant 0 if $\mathbb{T}=\mathbb{R}$ while it is constant 1 for $\mathbb{T}=\mathbb{Z}$. However, a time scale $\mathbb{T}$ could have nonconstant graininess.

A point $t\in\mathbb{T}$ is called right-scattered, right-dense, left-scattered and left-dense, if $\sigma(t)>t, \sigma(t)=t, \rho(t)<t$ and $\rho(t)=t$ holds, respectively. Throughout we let $a,b\in\mathbb{T}$ with $a<b$. For an interval $[a,b]\cap\mathbb{T}$ we simply write $[a,b]$ when this is not ambiguous. We also define

$$[a,b]^\kappa := [a,b]\setminus(\rho(b),b] \quad \text{and} \quad [a,b]^{\kappa^2} := [a,b]\setminus(\rho(\rho(b)),b].$$

We say that a function $f: \mathbb{T}\to\mathbb{R}$ is delta differentiable at $t\in\mathbb{T}^\kappa$ provided there exists a real number $f^\Delta(t)$ such that for all $\varepsilon>0$, there is a neighborhood $U=(t-\delta,t+\delta)\cap\mathbb{T}$ of $t$ with

$$\left|f(\sigma(t))-f(s)-f^\Delta(t)(\sigma(t)-s)\right|\leq\varepsilon|\sigma(t)-s| \quad \text{for all} \quad s\in U.$$

For differentiable $f$, the formula

$$f^\sigma = f + \mu f^\Delta \tag{2}$$

is very useful and easy to prove. If $f$ and $g$ are both differentiable, then so is $fg$ with

$$(fg)^\Delta = f^\Delta g + f^\sigma g^\Delta, \tag{3}$$

where we abbreviate $f\circ\sigma$ by $f^\sigma$.

Next, a function $f: \mathbb{T}\to\mathbb{R}$ is called rd-continuous if it is continuous in right-dense points and if its left-sided limits exist in left-dense points. By $C_{rd}$ we denote the set of all rd-continuous functions, while $C_{rd}^1$ denotes the set of all differentiable functions with rd-continuous derivative. It is known that rd-continuous functions possess an antiderivative, i.e., there exists a function $F: \mathbb{T}\to\mathbb{R}$ with $F^\Delta(t)=f(t)$, and in this case an integral of $f$ from $a$ to $b$ ($a,b \in \mathbb{T}$) is defined by

$$\int_a^b f(t)\Delta t = F(b) - F(a). \tag{4}$$

## 3. Variational relationships on time scales

*3.1 Exchange relationship between the isochronous variation and the delta derivatives*

Consider two infinitely closed orbits $\alpha$ and $\alpha+d\alpha$. We denote the generalized coordinates by $q=q(t,\alpha)$, $q=q(t,\alpha+d\alpha)$, corresponding to the two infinitely closed orbits respectively in giving time on time scale $\mathbb{T}$.

We define the isochronous variation as

$$\delta q = q(t, \alpha + d\alpha) - q(t, \alpha). \tag{5}$$

Extending $q=q(t,\alpha+d\alpha)$ to the linear terms of $d\alpha$, we obtain

$$q(t, \alpha + d\alpha) = q(t, \alpha) + \frac{\partial q(t,\alpha)}{\partial \alpha} d\alpha. \tag{6}$$

Substituting Eq. (6) into Eq. (5), we have

$$\delta q = \frac{\partial q(t,\alpha)}{\partial \alpha} d\alpha. \tag{7}$$

Similarly we have

$$\frac{\Delta}{\Delta t} \delta q = \frac{\Delta}{\Delta t} \frac{\partial q(t,\alpha)}{\partial \alpha} d\alpha. \tag{8}$$

According to Eq. (6), we get

$$\delta q^\Delta = q^\Delta(t, \alpha + d\alpha) - q^\Delta(t, \alpha) = \frac{\Delta}{\Delta t} \frac{\partial q(t,\alpha)}{\partial \alpha} d\alpha. \tag{9}$$

Comparing Eqs. (9) and (8), we obtain

$$\delta q^\Delta = (\delta q)^\Delta. \tag{10}$$

Similarly we have

$$\delta q^\sigma = (\delta q)^\sigma. \tag{11}$$

We call Eqs. (10) and (11) the exchanging relationships with respect to the delta derivatives and isochronous variation.

*3.2 The isochronous variation on time scales*

Now, we study the infinitely closed orbits $\alpha$ and $\alpha+d\alpha$. The generalized coordinates are given by $q=q(t,\alpha)$ and $q^*=q^*(t,\alpha)$ for any $t \in \mathbb{T}$, where $t=t(\alpha)$, so we have

$$q = q[t(\alpha), \alpha].$$

Taking total variation for $q$, we obtain

$$\Delta q = \frac{\partial q[t(\alpha),\alpha]}{\partial \alpha}d\alpha + \frac{\partial q[t(\alpha),\alpha]}{\partial t}\frac{\partial t}{\partial \alpha}d\alpha. \tag{12}$$

Since $\Delta t$ is the variational of time with respect to $\alpha$, therefore

$$\Delta t = \frac{\partial t}{\partial \alpha}d\alpha. \tag{13}$$

Substituting Eqs. (7) and (13) into Eq. (12), we get

$$\Delta q = \delta q + q^{\Delta}\Delta t. \tag{14}$$

We call Eq. (14) the relationship between the isochronous variation and the total variation on time scale $\mathbb{T}$.

Using Eqs. (10) and (14) we have

$$\Delta q^{\Delta} = \delta q^{\Delta} + q^{\Delta\Delta}\cdot\Delta t. \tag{15}$$

Differentiating both sides of Eq. (14) with respect to $t$, we obtain

$$(\Delta q)^{\Delta} = \delta q^{\Delta} + q^{\Delta\Delta}\cdot(\Delta t)^{\sigma} + q^{\Delta}\cdot(\Delta t)^{\Delta}. \tag{16}$$

According to Eqs. (15) and (16), we get

$$\Delta q^{\Delta} = (\Delta q)^{\Delta} - q^{\Delta\Delta}(\Delta t)^{\sigma} - q^{\Delta}(\Delta t)^{\Delta} + q^{\Delta\Delta}\cdot\Delta t = (\Delta q)^{\Delta} - q^{\Delta}(\Delta t)^{\Delta} - \mu(t)\cdot(\Delta t)^{\Delta}q^{\Delta\Delta}. \tag{17}$$

Similarly we have

$$\Delta q^{\Delta\Delta} = \delta q^{\Delta\Delta} + q^{\Delta\Delta\Delta}\cdot\Delta t, \tag{18}$$

$$(\Delta q)^{\Delta\Delta} = \delta q^{\Delta\Delta} + q^{\Delta\Delta\Delta}(\Delta t)^{\sigma^2} + q^{\Delta\Delta}(\Delta t)^{\sigma\Delta} + q^{\Delta\Delta}(\Delta t)^{\Delta\sigma} + q^{\Delta}(\Delta t)^{\Delta\Delta}. \tag{19}$$

From Eqs. (18) and (19) we have

$$\Delta q^{\Delta\Delta} = q^{\Delta\Delta\Delta}\cdot\Delta t + (\Delta q)^{\Delta\Delta} - q^{\Delta\Delta\Delta}(\Delta t)^{\sigma^2} - q^{\Delta\Delta}(\Delta t)^{\sigma\Delta} - q^{\Delta\Delta}(\Delta t)^{\Delta\sigma} - q^{\Delta}(\Delta t)^{\Delta\Delta}. \tag{20}$$

Differentiating both sides of Eq. (17) with respect to $t$, we obtain

$$\begin{aligned}(\Delta q^{\Delta})^{\Delta} = & (\Delta q)^{\Delta\Delta} - q^{\Delta\Delta\Delta}(\Delta t)^{\sigma^2} - q^{\Delta\Delta}(\Delta t)^{\sigma\Delta} - q^{\Delta\Delta}(\Delta t)^{\Delta\sigma} - q^{\Delta}(\Delta t)^{\Delta\Delta} \\ & + q^{\Delta\Delta\Delta}(\Delta t)^{\sigma} + q^{\Delta\Delta}(\Delta t)^{\Delta}.\end{aligned} \tag{21}$$

By virtue of Eqs. (20) and (21) we can obtain

$$\Delta q^{\Delta\Delta} = (\Delta q^{\Delta})^{\Delta} - q^{\Delta\Delta}(\Delta t)^{\Delta} - \mu(t)\cdot(\Delta t)^{\Delta}q^{\Delta\Delta\Delta}. \tag{22}$$

**4. Lie symmetries of Lagrangian systems on time scales**

*4.1 Equations of motion of the systems on time scales*

We consider the fundamental problem of the calculus of variations on time scales as defined by Bohner [2]:

$$I[q(\cdot)] = \int_a^b L(t, q^\sigma(t), q^\Delta(t)) \Delta t \to \min, \quad (q(a) = \alpha) \quad (q(b) = \beta) \tag{23}$$

under given boundary conditions $q(a) = A$, $q(b) = B$, where σ is the forward jump operator and $q^\Delta$ is the delta derivative of $q$ with respect to $\mathbb{T}$, and the Lagrangian $L : \mathbb{R} \times \mathbb{R}^n \times \mathbb{R}^n \to \mathbb{R}$ is a $C^1$ function with respect to its arguments. By $\partial_i L$ we will denote the partial derivative of $L$ with respect to the $i$th variable, $i$=1, 2, 3. Admissible function $q(\cdot)$ are assumed to be $C^1_{rd}$.

The following results, known as the Euler-Lagrange equation and the DuBois-Reymond equation, are necessary optimality conditions for optimal trajectories of delta variational problems.

**Theorem 1** (Euler-Lagrange Equation [2]). If $q(\cdot)$ is a minimizer of problem (23), then $q(\cdot)$ satisfies the equation

$$\frac{\Delta}{\Delta t} \partial_3 L(t, q^\sigma(t), q^\Delta(t)) = \partial_2 L(t, q^\sigma(t), q^\Delta(t)). \tag{24}$$

**Theorem 2** (DuBois-Reymond Equation for Delta Problems [4]). If $q(\cdot) \in C^1_{rd}$ is a local minimizer of problem (23), then $q(\cdot)$ satisfies the equation

$$\frac{\Delta}{\Delta t}\left[-L(t, q^\sigma, q^\Delta) + \partial_3 L(t, q^\sigma, q^\Delta) \cdot q^\Delta + \partial_1 L(t, q^\sigma, q^\Delta) \cdot \mu(t)\right] \\ = -\partial_1 L(t, q^\sigma(t), q^\Delta(t)). \tag{25}$$

We observe that the mechanical system with Lagrangian $L(t, q^\sigma, q^\Delta)$ has the equation of motion

$$\frac{\Delta}{\Delta t} \partial_3 L(t, q^\sigma(t), q^\Delta(t)) - \partial_2 L(t, q^\sigma(t), q^\Delta(t)) = 0. \tag{26}$$

In general, it is supposed that the system (26) is nonsingular, i.e.

$$\frac{\partial^2 L}{\partial (q^\Delta)^2} + \mu(1 + \mu^\Delta) \frac{\partial^2 L}{\partial q^\Delta \partial q^\sigma} \neq 0. \tag{27}$$

Expanding Eq. (26), we can determine the generalized acceleration as

$$q^{\Delta\Delta} = h(t, q^\sigma, q^\Delta). \tag{28}$$

*4.2 Infinitesimal transformations and determining equations on time scales*

Introduce the infinitesimal transformations in terms of time and coordinates

$$t^* = t + \Delta t, \quad q^*(t) = q(t) + \Delta q, \tag{29}$$

or their expanded form

$$t^* = t + \varepsilon\tau(t,q) + o(\varepsilon), \quad q^*(t) = q(t) + \varepsilon\xi(t,q) + o(\varepsilon), \tag{30}$$

where $\varepsilon$ is an infinitesimal parameter, and $\tau(t,q)$, $\xi(t,q)$ are the generators of infinitesimal transformations. Eq. (30) is a one-parameter Lie-point group of transformations. Introducing the vector's generator under the infinitesimal transformations

$$X^{(0)} = \tau\frac{\partial}{\partial t} + \xi^\sigma\frac{\partial}{\partial q^\sigma}, \tag{31}$$

which can be prolonged to the two- and three-point schemes

$$X^{(1)} = \tau\frac{\partial}{\partial t} + \xi^\sigma\frac{\partial}{\partial q^\sigma} + \left(\xi^\Delta - \tau^\Delta q^{\Delta\sigma}\right)\frac{\partial}{\partial q^\Delta}, \tag{32}$$

$$X^{(2)} = X^{(1)} + \left[\left(\xi^\Delta - \tau^\Delta q^{\Delta\sigma}\right)^\Delta - \tau^\Delta q^{\Delta\Delta\sigma}\right]\frac{\partial}{\partial q^{\Delta\Delta}}, \tag{33}$$

where

$$\xi^\sigma(t,q(t)) = \xi(\sigma(t), q(\sigma(t))), \quad \xi^\Delta(t,q(t)) = \frac{\Delta}{\Delta t}\xi(t,q(t)),$$

and $\tau(t,q)$, $\xi(t,q)$ are infinitesimal generators. Then based on the invariance of the differential Eqs. (28) under the infinitesimal transformations (30), if and only if

$$X^{(2)}\left[q^{\Delta\Delta} - h(t, q^\sigma, q^\Delta)\right] = 0, \tag{34}$$

we can have

$$\xi^{\Delta\Delta} - \tau^{\Delta\Delta}q^\Delta - \left[(\mu\tau^\Delta)^\Delta + 2\tau^\Delta + \mu\tau^{\Delta\Delta}\right]h - \left[\mu(\mu\tau^\Delta)^\Delta + 2\mu\tau^\Delta\right]q^{\Delta\Delta\Delta} = X^{(1)}h. \tag{35}$$

Eqs. (35) are called the determining equations on time scales $\mathbb{T}$ which the generators $\tau(t,q)$ and $\xi(t,q)$ should satisfy.

We hereinafter give the definition of Lie symmetries of the Lagrangian systems on time scales.

**Definition 1**. If generators $\tau(t,q)$, $\xi(t,q)$ satisfy the determining equations (35), then corresponding symmetries are called Lie symmetries of Lagrangian systems on time scales (26).

*4.3 Lie symmetrical structural equations and conserved quantities on time scales*

Lie symmetries don't always generate conserved quantities. The subsequent propositions give the condition under which Lie symmetries generate conserved quantities and the form of conserved quantities with delta derivatives.

**Definition 2.** Quantity $I(t,q,q^\sigma,q^\Delta)$ is said to be a conserved quantity if and only if $\frac{\Delta}{\Delta t}I(t,q,q^\sigma(t),q^\Delta(t))=0$ is preserved along all $q(t)$ that satisfy the Euler-Lagrange equation (26).

**Theorem 3.** For the infinitesimal generators $\tau$ and $\xi$ satisfying the determining Eq. (35), if there is a gauge function $G=G(t,q^\sigma,q^\Delta)$ satisfying the following equation

$$L\tau^\Delta + X^{(1)}L + \partial_3 L \cdot \mu\tau^\Delta q^{\Delta\Delta} + G^\Delta = 0, \qquad (36)$$

then the system possesses a conserved quantity with delta derivatives

$$I(t,q^\sigma,q^\Delta) = \partial_3 L(t,q^\sigma,q^\Delta)\cdot \xi(t,q)$$

$$+ [L(t,q^\sigma,q^\Delta) - \partial_3 L(t,q^\sigma,q^\Delta)\cdot q^\Delta - \partial_1 L(t,q^\sigma,q^\Delta)\cdot \mu(t)]\cdot \tau(t,q) = \text{const.} \qquad (37)$$

**Proof.** Using the Euler-Lagrange equation (24), the DuBois-Reymond equation (25) and the structure equation (36), we obtain

$$\frac{\Delta}{\Delta t}[\partial_3 L(t,q^\sigma,q^\Delta)\cdot \xi(t,q) + (L(t,q^\sigma,q^\Delta) - \partial_3 L(t,q^\sigma,q^\Delta)\cdot q^\Delta - \partial_1 L(t,q^\sigma,q^\Delta)\cdot \mu(t))\cdot \tau(t,q)$$

$$+ G(t,q^\sigma,q^\Delta)] = \frac{\Delta}{\Delta t}\partial_3 L(t,q^\sigma,q^\Delta)\cdot \xi^\sigma(t,q) + \partial_3 L(t,q^\sigma,q^\Delta)\cdot \xi^\Delta(t,q)$$

$$+ \tau^\sigma(t,q)\cdot \frac{\Delta}{\Delta t}(L(t,q^\sigma,q^\Delta) - \partial_3 L(t,q^\sigma,q^\Delta)\cdot q^\Delta - \partial_1 L(t,q^\sigma,q^\Delta)\cdot \mu(t))$$

$$+ \tau^\Delta(t,q)\cdot (L(t,q^\sigma,q^\Delta) - \partial_3 L(t,q^\sigma,q^\Delta)\cdot q^\Delta - \partial_1 L(t,q^\sigma,q^\Delta)\mu(t)) + G^\Delta(t,q^\sigma,q^\Delta)$$

$$= \partial_2 L(t,q^\sigma,q^\Delta)\cdot \xi^\sigma(t,q) + \partial_3 L(t,q^\sigma,q^\Delta)\cdot \xi^\Delta(t,q) + \partial_1 L(t,q^\sigma,q^\Delta)\cdot \tau(t,q)$$

$$+ L(t,q^\sigma,q^\Delta)\cdot \tau^\Delta(t,q) - \partial_3 L(t,q^\sigma,q^\Delta)\cdot \tau^\Delta(t,q)\cdot q^\Delta + G^\Delta(t,q^\sigma,q^\Delta)$$

$$= L(t,q^\sigma,q^\Delta)\tau^\Delta + X^{(1)}L(t,q^\sigma,q^\Delta) + \partial_3 L(t,q^\sigma,q^\Delta)\cdot \mu\tau^\Delta q^{\Delta\Delta} + \frac{\Delta}{\Delta t}G(t,q^\sigma,q^\Delta) = 0.$$

Eq. (36) is called the structure equation for the Lagrangian systems on time scales.

If the transformations (30) satisfy Noether's identity

$$\partial_1 L(t,q^\sigma,q^\Delta) \cdot \tau(t,q) + \partial_2 L(t,q^\sigma,q^\Delta) \cdot \xi^\sigma(t,q) + \partial_3 L(t,q^\sigma,q^\Delta) \cdot \xi^\Delta(t,q)$$
$$+ L(t,q^\sigma,q^\Delta) \cdot \tau^\Delta - \partial_3 L(t,q^\sigma q^\Delta) \cdot \tau^\Delta \cdot q^\Delta = -\frac{\Delta}{\Delta t} G(t,q^\sigma,q^\Delta), \quad (38)$$

then the transformations (30) are called the Noether symmetrical transformations of the system (26). We have:

**Theorem 4.** The structure equation with delta derivatives (36) of the Lie symmetry is equivalent to Noether's identity (38).

The method of solution of the direct problem of the Lie symmetries on time scales is the following: firstly, establish the determining equation with delta derivatives (35), and seek the generator $\tau, \xi$ from these equations; secondly, substitute the generator obtained into the structure equation (36) to determine $G(t,q^\sigma,q^\Delta)$; finally, substitute $\tau, \xi$ and $G(t,q^\sigma,q^\Delta)$ into the formula (37) to get the conserved quantities of the symmetries on time scales.

## 5. Discussion

In the continuous time ($\mathbb{T}=\mathbb{R}$), previous results are reduced to the classical results of the Lie symmetries of the Lagrangian systems [24]. Here we will consider the Lie symmetries of discrete Lagrangian systems on discrete-time ($\mathbb{T}=\mathbb{Z}$). The time $t_k=k$ is a discrete variable: $k\in\mathbb{Z}$. The horizon consists of $N$ periods, $t_k=k=M,M+1,...,M+N-1$ where $M$ and $N$ are fixed integers, instead of a continuous interval. The purpose of the following work is to deduce the discrete equations of motion, the discrete structure equations and the discrete conserved quantities on discrete-time $\mathbb{T}=\mathbb{Z}$ from the results we have obtained on an arbitrary time scale $\mathbb{T}$. We refer to the literatures for further reading on Lie symmetries of the mechanical systems for the discrete case [29].

When $\mathbb{T}=\mathbb{Z}$, problem (23) is reduced to a one-freedom, discrete calculus of variations in which the fundamental problem is to select among all finite sequence $\{q_{k+1}\}$, the one which minimizes the sum

$$I[q_{k+1}] = \sum_{k=M}^{M+N-1} L(t,q_{k+1},\Delta q), \quad (39)$$

where $\Delta$ is the difference operator, $\Delta q = q_{k+1} - q_k$.

Eqs. (24), (25) can be written in the form of discrete Euler-Lagrange equations:

$$\frac{\partial L(k, q_{k+1}, \Delta q)}{\partial (\Delta q)} - \frac{\partial L(k-1, q_k, \Delta_1 q)}{\partial (\Delta_1 q)} - \frac{\partial L(k, q_{k+1}, \Delta q)}{\partial q_{k+1}} = 0, \quad (40)$$

$$\Delta \left[ \frac{\partial L(k, q_{k+1}, \Delta q)}{\partial (\Delta q)} \cdot \Delta q + \frac{\partial L(k, q_{k+1}, \Delta q)}{\partial k} - L(k, q_{k+1}, \Delta q) \right] = -\frac{\partial L(k, q_{k+1}, \Delta q)}{\partial k}, \quad (41)$$

$$(k = M, M+1, \ldots, M+N-1),$$

where $\Delta_1 q = q_k - q_{k-1}$.

The discrete equations of motion can be put in the form

$$\Delta^2 q = h(k, q_{k+1}, \Delta q) \quad (42)$$

where we abbreviate $\Delta(\Delta q)$ by $\Delta^2 q$.

The vector field of generators (31) turns out to be

$$X_*^{(0)} = \tau_k \frac{\partial}{\partial t} + \xi_{k+1} \frac{\partial}{\partial q_{k+1}}, \quad (43)$$

which can be prolonged to the two- and three-point schemes

$$X_*^{(1)} = \tau_k \frac{\partial}{\partial t} + \xi_{k+1} \frac{\partial}{\partial q_{k+1}} + \frac{\partial}{\partial (\Delta q)} \left( \Delta \xi - \Delta \tau \cdot \Delta q - \Delta \tau \cdot \Delta^2 q \right) \quad (44)$$

$$X_*^{(2)} = X_*^{(1)} + \left[ \Delta \left( \Delta \xi - \Delta \tau \cdot \Delta q - \Delta \tau \cdot \Delta^2 q \right) - \Delta \tau \cdot \Delta^2 q - \Delta \tau \cdot \Delta^3 q \right] \frac{\partial}{\partial (\Delta^2 q)}. \quad (45)$$

The invariance of discrete Euler-Lagrange equations (40) under the infinitesimal transformations (30) leads to the satisfaction of the following discrete determining equations:

$$\Delta^2 \xi - \Delta q \cdot \Delta^2 \tau - 2(\Delta \tau + \Delta^2 \tau) \Delta^2 q - (2\Delta \tau + \Delta^2 \tau) \Delta^3 q = X_*^{(1)}(\Delta^2 q), \quad (46)$$

where $\Delta \xi = \xi_{k+1} - \xi_k$, $\Delta \tau = \tau_{k+1} - \tau_k$.

We hereinafter give the Lie symmetries of the discrete Lagrangian systems on discrete-time ($\mathbb{T} = \mathbb{Z}$).

**Corollary 1.** If generator $\tau_k, \xi_k$ satisfies the discrete determining equations (46), then corresponding symmetries are called Lie symmetries of the discrete Lagrangian systems (42).

**Corollary 2.** If the infinitesimal generators $\tau_k, \xi_k$ satisfy Eq. (46), and in addition there exists a discrete gauge function $G = G(k, q_{k+1}, \Delta q)$ such that the identity

$$L(k,q_{k+1},\Delta q)\cdot \Delta \tau + X_*^{(1)}L(k,q_{k+1},\Delta q) + \partial_3 L(k,q_{k+1},\Delta q)\cdot \Delta \tau \cdot \Delta^2 q \\ + \Delta G(k,q_{k+1},\Delta q) = 0 \quad (47)$$

holds, then the discrete Lagrangian systems possesses the discrete conserved quantities

$$I(k,q_k,q_{k+1},\Delta q) = \partial_3 L(k,q_{k+1},\Delta q)\cdot \xi_k(k,q) + [L(k,q_{k+1},\Delta q)$$
$$- \partial_3 L(k,q_{k+1},\Delta q)\Delta q - \partial_1 L(k,q_{k+1},\Delta q)]\tau_k(k,q) = \text{const.} \quad (48)$$

**Proof.** Using the discrete Euler-Lagrange equations (40), (41), and the discrete structure equation (47), we obtain

$$\Delta I = \frac{\partial L(k,q_{k+1},\Delta q)}{\partial q_{k+1}}\xi_{k+1} + \frac{\partial L(k,q_{k+1},\Delta q)}{\partial(\Delta q)}\cdot \Delta \xi + \frac{\partial L(k,q_{k+1},\Delta q)}{\partial k}\tau_{k+1}$$
$$+ \left(L(k,q_{k+1},\Delta q) - \frac{\partial L(k,q_{k+1},\Delta q)}{\partial(\Delta q)}\cdot \Delta q - \frac{\partial L(k,q_{k+1},\Delta q)}{\partial k}\right)\Delta \tau$$
$$= \frac{\partial L(k,q_{k+1},\Delta q)}{\partial k}\tau_k + \frac{\partial L(k,q_{k+1},\Delta q)}{\partial q_{k+1}}\xi_{k+1} + \frac{\partial L(k,q_{k+1},\Delta q)}{\partial(\Delta q)}\cdot \Delta \xi$$
$$+ \left[L(k,q_{k+1},\Delta q) - \frac{\partial L(k,q_{k+1},\Delta q)}{\partial(\Delta q)}\Delta q\right]\Delta \tau + \Delta G(k,q_{k+1},\Delta q) = 0$$

whence Eq. (48) holds. This completes the proof.

Eq. (47) is called the discrete structure equation corresponding to the Lie symmetries of discrete Lagrangian systems, and Eq. (48) is called discrete conserved quantities associated with the systems.

Furthermore, if the transformations (30) satisfy Noether's identity

$$\frac{\partial L}{\partial k}\tau_k + \frac{\partial L}{\partial q_{k+1}}\xi_{k+1} + \frac{\partial L}{\partial(\Delta q)}\cdot \Delta \xi + \left[L - \frac{\partial L}{\partial(\Delta q)}\Delta q\right]\Delta \tau + \Delta G = 0, \quad (49)$$

then the discrete transformations (30) are called the discrete Noether symmetrical transformations of the system (42). We also have:

**Corollary 3.** The discrete structure equation (47) of the Lie symmetry is equivalent to the discrete Noether identity (49).

## 6. Example

**Example 1** We first consider an example of a conservation law of a Lagrangian system on a discrete but nonhomogeneous time scale (graininess is not conatant). The time scale and the Lagrangian of the system are

$$\mathbb{T} = \{2^n : n \in \mathbb{N} \cup \{0\}\}, \tag{50}$$

and

$$L(t, q^\sigma, q^\Delta) = t + q^\sigma q^\Delta \tag{51}$$

Equation (28) gives the equation of motion of the system

$$q^{\Delta\Delta} = -\frac{q^\Delta}{2t}. \tag{52}$$

The determining equation of the Lie symmetries of Eq. (52) under the infinitesimal transformation $\xi=\xi(t,q)$, $\tau=\tau(t,q)$ is

$$\xi^{\Delta\Delta} + \frac{\xi^\Delta}{2t} = \tau \cdot \frac{q^\Delta}{4t^2} + \frac{\tau^{\Delta\Delta} q^\Delta}{4} - \frac{\tau^\Delta q^\Delta}{8t}. \tag{53}$$

We can obtain the following solution of Eq. (53):

$$\xi = \frac{\ln t}{\ln 2}, \quad \tau = 0. \tag{54}$$

The structure equation with delta derivatives (36) gives

$$\tau^\Delta t + \tau + q^\Delta \xi^\sigma + q^\sigma \xi^\Delta = -G^\Delta. \tag{55}$$

Substituting the generators (54) into the structure equation (55) yields

$$G = -\frac{q \ln 2t}{\ln 2}. \tag{56}$$

According to Theorem 3, substituting the generator (54) and the gauge function (56) into the formula (37), we get the following conserved quantity

$$I(t, q, q^\sigma, q^\Delta) = \frac{t q^\Delta \ln t}{\ln 2} - q = \text{const.}$$

In this example, the transformation is also Noether symmetrical and such fact is easily verified by direct application of Definition 2: $\frac{\Delta I}{\Delta t} = (t\xi)^\Delta q^{\Delta\sigma} + t\xi q^{\Delta\Delta} - q^\Delta = t\xi q^{\Delta\Delta} - q^\Delta$
$+ (\xi^\sigma + t\xi^\Delta)(q^\Delta + tq^{\Delta\Delta}) = t\xi^\Delta q^\Delta - q^\Delta = 0$.

## 7. Conclusion

In this work the Lie symmetries of Lagrangian systems on an arbitrary time scale $\mathbb{T}$ are investigated. This is a significant work which unifies and extends the previous formulations of Lie's method in the discrete-time and continuous domains.

We have obtained the condition under which Lie symmetry can lead to a conserved quantity, and we have also found the corresponding conserved quantities with delta derivatives from a known Lie symmetry. The results indicate that it is also a promising approach to seek the Lie symmetries of discrete Lagrangian systems on discrete-time ($\mathbb{T}=\mathbb{Z}$). Using this approach, it might also be possible to obtain Lie symmetries of the nonconservative and the nonholonomic mechanical systems with delta derivatives.

## Acknowledgments

The authors would like to express their sincere thanks to referee for the valuable advice. This work is supported by the Natural Science Foundation of China (Grant No. 11072218).

## References

[1] B. Aulbach, S. Hilger, Qualitative Theory of Differential Equations , Szeged, 1988 .
[2] M. Bohner, Calculus of variations on time scales, Dynam. Syst. Appl. 13 (2004) 339-349.
[3] R. Hilscher, V. Zeidan, Calculus of variations on time scales, J. Math. Anal. Appl. 289 (2004) 143-166.
[4] N. Martins, D. F. M. Torres, Noether's symmetry theorem for nabla problems of the calculus of variations, Appl. Math. Lett. 23 (2010) 1432-1438.
[5] R. Almeida, D. F. M. Torres, Isoperimetric problems on time scales with nabla derivatives, J. Vib. Control 6 (2009) 951-958.
[6] A. B. Malinowska, D. F. M. Torres, Necessary and sufficient conditions for local Pareto optimality on time scales, J. Math. Sci. 6 (2009) 803-810.
[7] R.P. Agarwal, M. Bohner, Basic calculus on time scales and some of its applications, Results Math. 35 (1999) 3–22.
[8] Z. Bartosiewicz, E. Pawluszewicz, Realizations of nonlinear control systems on time scales, IEEE. T. Automat. Contr. 53 (2008) 571-575.
[9] M. Bohner, G. S. Guseinov, Double integral calculus of variations on time scales,

Computer Mathematics appl. 54 (2007) 45-57.

[10] R. Hilscher,V. Zeidan, Weak maximum principle and accessory problem for control problems on time scales, Nonlinear Anal. 70 (2009) 3209-3266.

[11] G. S. Guseinov, Integration on time scales, J. Math. Anal. Appl. 285 (2003) 107–127.

[12] R. P. Agarwala , M. Bohnerb, D. O'Reganc, Time scale boundary value problems on infinite intervals, J. Comput. Appl. Math. 141 (2002) 27–34.

[13] F. M. Atici, G.S. Guseinov, On Green's functions and positive solutions for boundary value problems on time scales, J. Comput. Appl. Math. 141 (2002) 75-99.

[14] F. M. Atici, D. C. Biles, A. Lebedinsky, An application of time scales to economics, Math. Comput. Model. 43 (2006) 718-726.

[15] Z. Bartosiewicz, D. F. M. Torres, Noether's theorem on time scales, J. Math. Anal. Appl. 342 (2008) 1220-1226.

[16] E. Noether, Invarianten beliebiger Differentialausdrücke, Nachr. Ges. Wiss. Göttingen, 57 (1918) 235-246.

[17] J. L. Fu, X. W. Li , C. R. Li , W. J. Zhao, B. Y. Chen, Symmetries and exact solutions of discrete nonconservative systems, Sci. China 53 (2010) 1699-1706.

[18] J. L. Fu, B. Y. Chen, L. Q. Chen, Noether symmetries of discrete nonholonomic dynamical systems, Phys. Lett. A 373 (2009) 409-412.

[19] S. Zhou , H. Fu, J. L. Fu, Symmetry theories of Hamiltonian systems with fractional derivatives, Sci. China 54 (2011) 1847-1853.

[20] R. K. Gazizov, N. H. Ibragimov, Lie Symmetry Analysis of Differential Equations in Finance, Nonlinear dynam. 17(1998) 387-407.

[21] J. L. Fu, B. Y. Chen, H. Fu, G. L.Zhao, R. W. Liu and Z. Y. Zhu,Velocity-dependent symmetries and non-Noether conserved quantities of electromechanical systems, Sci. China 54 (2011) 288-295.

[22] J. R. Gregory, Finding abstract Lie symmetry algebras of differential equations without integrating determining equations, Eur. J. Math. 2 (1991) 319-340.

[23] B. B. Juan , V. M. Pe´rez-Garcı´a, V. Vekslerchik, Lie Symmetries and Solitons in Nonlinear Systems with Spatially Inhomogeneous Nonlinearities, Phys. Rev. Lett. 98 (2007) 064102 .

[24] M. Lutzky, Dynamical symmetries and conserved quantities, J. Phys A: Math.Gen. 12(1979) 973-981.

[25] J. L. Fu, L. Q. Chen, B. Y. Chen, Noether-type theory for discrete mechanicoelectrical dynamical systems with nonregular lattices, Sci. China 53 (2010) 1687-1698.

[26] G. D. Matteis, L. Martina, Lie point symmetries and reductions of one-dimensional equations describing perfect Korteweg-type nematic fluids, J. Math. Phys. 53 (2012) 033101.

[27] H. B. Zhang, Lie symmetries and conserved quantities of non-holonomic mechanical systems with unilateral Vacco constraints, Chin. Phys. 11 (2002) 1-4.

[28] J. L. Fu, H. Fu, R. W. Liu, Hojman conserved quantities of discrete mechanico–electrical systems constructed by continuous symmetries, Phys. Lett. A 374 (2010) 1812-1818.

[29] H. B.Zhang, L. Q. Chen and R. W. Liu, Discrete variational principle and the first integrals of the conservative holonomic systems in event space, Chin. Phys.14 (2005) 888-892.